\documentclass[fleqn,usenatbib]{mnras}

\usepackage{newtxtext,newtxmath}
\usepackage[T1]{fontenc}

\DeclareRobustCommand{\VAN}[3]{#2}
\let\VANthebibliography\thebibliography
\def\thebibliography{\DeclareRobustCommand{\VAN}[3]{##3}\VANthebibliography}

\usepackage{graphicx}	% Including figure files
\usepackage{amsmath}	% Advanced maths commands
\newcommand{\vect}[1]{\boldsymbol{#1}}
\usepackage{comment}
\usepackage{lineno}
\usepackage{textgreek}
\usepackage{hyperref}

\newcommand{\rd}{r_\mathrm{d}}

\newcommand{\xis}{\xi_\mathrm{s}}
\newcommand{\xiw}{\xi_\mathrm{w}}
\newcommand{\Ps}{P_\mathrm{s}}
\newcommand{\Pw}{P_\mathrm{w}}
\newcommand{\Bs}{B_\mathrm{s}}
\newcommand{\Bw}{B_\mathrm{w}}
\newcommand{\Dv}{D_\mathrm{v}}
\newcommand{\glam}{\texttt{GLAM}}
\newcommand{\camb}{\texttt{CAMB}}
\newcommand{\emcee}{\texttt{emcee}}
\newcommand{\Mpc}{\mathrm{Mpc}}
\newcommand{\Plin}{P_\mathrm{lin}}
\newcommand{\alphav}{\alpha_\mathrm{V}}

\title[Modeling BAO in bispectrum]{Modeling the BAO feature in Bispectrum}

\author[J.~Behera et al.]{
Jayashree Behera$^{1}$\thanks{E-mail: \textcolor{blue}{jayashreeb@phys.ksu.edu}},
Mehdi Rezaie$^{1}$,
Lado Samushia$^{1,2,3}$,
and Julia Ereza$^{4}$
\\
$^{1}$Department of Physics, Kansas State University, 116 Cardwell Hall, Manhattan, KS, 66506, USA \\
$^{2}$E.Kharadze Georgian National Astrophysical Observatory, 47/57 Kostava St.,Tbilisi 0179, Georgia\\
$^{3}$School of Natural Sciences and Medicine, Ilia State University, 3/5 Cholokashvili Ave., Tbilisi 0162, Georgia\\
$^{4}$Instituto de Astrof\'isica de Andaluc\'ia (CSIC), Glorieta de la Astronom\'ia, E-18080 Granada, Spain\\
}

\pubyear{2023}

\nolinenumbers

\begin{document}
\label{firstpage}
\pagerange{\pageref{firstpage}--\pageref{lastpage}}
\maketitle

\begin{abstract}
We investigate how well a simple leading order perturbation theory model of the bispectrum can fit the BAO feature in the measured bispectrum monopole of galaxies. Previous works showed that perturbative models of the galaxy bispectrum start failing at the wavenumbers of $k \sim 0.1 h\ \Mpc^{-1}$. We show that when the BAO feature in the bispectrum is separated it can be successfully modeled up to much higher wavenumbers. We validate our modeling on \texttt{GLAM} simulations that were run with and without the BAO feature in the initial conditions. We also quantify the amount of systematic error due to BAO template being offset from the true cosmology. We find that the systematic errors do not exceed 0.3 per cent for reasonable deviations of up to 3 per cent from the true value of the sound horizon.

\end{abstract}

%%%%%%%%%%%%%%%%%%%%%%%%%%%%%%%%%%%%%%%%%%%%%%%%%%

%%%%%%%%%%%%%%%%% BODY OF PAPER %%%%%%%%%%%%%%%%%%

\section{Introduction}

The Baryon Acoustic Oscillation (BAO) signal is a statistical feature imprinted into the spatial distribution of galaxies. Small overdensities in the initial distribution of dark matter serve as seeds for acoustic wave propagation in the early baryon-photon fluid. The propagation of these waves freezes at later times and results in a slightly overdense spherical shell of a radius of about $\rd \sim 100h^{-1}$ Mpc around each primordial overdensity. From the observational point of view, this means that every galaxy created around the initial overdensity has a slightly higher chance of having a neighboring galaxy on this spherical shell. Later this feature manifests itself as a local maximum (peak) at $\rd$ in the two-point correlation function of galaxies and as a decaying oscillatory feature of frequency $2\pi/\rd$ in the power spectrum. The BAO scale has been measured from the two-point correlation function with a sub-percent precision. Theoretical modeling and interpretation of the measured BAO scale in the two-point correlation function are relatively straightforward. The measurements in conjunction with the theory can be used to determine the distance-redshift relationship and therefore to constrain cosmological models. The same feature must also be present in other measures of galaxy clustering but it may not be as easy to model. The main goal of our paper is to explore the possibility of modeling the BAO feature in the Fourier space measurements of the galaxy three-point correlation function. 

Many methods for measuring the BAO peak position (or frequency) have been proposed in recent years. They all follow the same schematic idea proposed in \cite{2007ApJ...664..660E}. 

\begin{itemize}
\item Theoretically computed two-point correlation function (or the power spectrum) is split into ``smooth'' and ``wiggly'' components, $\xi = \xis\xiw$ or $P = \Ps\Pw$. This can be done in many different ways but the final result is not sensitive to details of the decomposition. With proper decomposition, $\Pw$ ends up being a decaying oscillation around unity \citep{2015PhRvD..92d3514B, 2015JCAP...09..014V, 2017JCAP...11..039F}.

\item The wiggly component is separated from the measurements in a similar way.
\item The leading order effect of non-linear evolution on the wiggly part of the power spectrum is scale-dependent damping of the oscillations, $\Pw \rightarrow (\Pw - 1)D(k) + 1$. The simplest model $D(k) = \exp(-\Sigma^2k^2)$, where $\Sigma$ is a nuisance parameter, works reasonably well \citep{2008JCAP...10..031N,2012MNRAS.427.2537C,2016MNRAS.458..613P,2016MNRAS.460.2453S,2018MNRAS.476..814H,2019JCAP...09..017C,2020MNRAS.493.4078H,2020PhRvD.102d3530V,2021MNRAS.501.2862S,2022JCAP...08..007B}. 
\item The damped model is supplemented by nuisance parameters to account for small inaccuracies in separating the smooth part. These nuisance parameters are usually polynomials in $k$ and they are smooth in the sense that they have little effect on the frequency of the oscillatory signal \citep{2017MNRAS.464.1168R,2018MNRAS.473.4773A,2021MNRAS.500.1201H,2021MNRAS.500.3254R}.
\item The model is then dilated $\Pw \rightarrow A\Pw(\alpha k)$ to align its oscillations with the ones in the measurements.
\end{itemize}

In the simplest version of this analysis (when the fit is performed on the angularly averaged power spectrum), the dilation parameter $\alpha$ depends on the distance redshift relationship,

\begin{equation} \label{alphadefinition}
\alpha \equiv \alphav = \frac{D_\mathrm{v}(z) \rd^\mathrm{tmp}}{D^\mathrm{fid}_\mathrm{v}(z) \rd} ,
\end{equation}
where $\Dv$ is the distance to the galaxies, 
\begin{equation} \label{eu_eqn}
D_\mathrm{v}(z) = \Bigg[\frac{cz D_\mathrm{A}(z)^2}{H(z)}\Bigg]^\frac{1}{3} ,
\end{equation}
$D_\mathrm{A}$ is the angular diameter distance, and $H$ is the Hubble parameter. The superscript $\mathrm{fid}$ denotes values in ``fiducial'' cosmology assumed when converting redshifts of galaxies to distances and the superscript $\mathrm{tmp}$ is used to denote the BAO scale in the BAO template which may be (but does not have to be) identical to the fiducial cosmology. Quantities without superscripts are the real values corresponding to the data. The fiducial values assumed in the theoretical model and the BAO peak position in the template are known exactly. Thus, measuring $\alpha$ allows us to infer the ratio of the distance at a given redshift to the BAO scale, a quantity that is extremely sensitive to the properties of dark energy. The exact procedure used in high-precision data analysis is more involved and requires great care in every small detail. The fit is performed to both the angular monopole and the quadrupole of the correlation function, which enables us to also measure the anisotropic dilation parameter $\epsilon = H^\mathrm{fid}\rd^\mathrm{tmp}/(H\rd)$ \citep{2008PhRvD..77l3540P,2012MNRAS.426.2719R,2014JCAP...04..001B,2015MNRAS.451.1331R,2017MNRAS.466.2242B,2018MNRAS.480.1096Z,2018MNRAS.480.2521H,2021MNRAS.500.1201H}, but the general idea remains the same.
These scalings have been shown to be robust for a wide variety of cosmological models \citep{2013arXiv1312.4996V,2015JCAP...01..034T,2018MNRAS.479.1021D,2014MNRAS.445....2V}.

This procedure for extracting the BAO signal from the power spectrum has been successfully used on many galaxy samples \citep{2020MNRAS.499..210N, 2021MNRAS.500..736B, 2021MNRAS.501.5616D, 2020MNRAS.498.2492G, 2021MNRAS.500.3254R}. The resulting distance measurements are leading sources of dark energy parameter constraints \citep{2014A&A...568A..22B,2015PhRvD..92l3516A,2017MNRAS.470.2617A,2017MNRAS.467.2085G,2017MNRAS.464.1640S,2018PhRvD..98d3526A,2020PhRvD.101l3516A,2020JCAP...05..005D,2020PhRvD.101f3502D,2020JCAP...05..042I,2020A&A...641A...6P,2020A&A...633L..10T, 2023arXiv230714802E}. 

The physics governing the production of Baryon Acoustic Oscillations (BAOs) in the matter power spectrum is well understood \citep{1968ApJ...151..459S,1970ApJ...162..815P,1970Ap&SS...7....3S,1984ApJ...285L..45B,1987MNRAS.226..655B, 1989ApJS...71....1H}. 
In the very early universe, the matter distribution was highly uniform. Small overdensities in this uniform distribution were growing in time with gravity, forming dark matter halos and later galaxies. Pulling matter into overdensities raised the temperature of matter, thereby creating an outward radiation pressure that pushed the matter apart again. As it expanded though, it cooled and gravity started to pull it back again. This interactivity between gravity and pressure set up an oscillation that created the equivalent of spherical sound waves that spread outward like bubbles. When the Universe reached the age of around 380,000 years, atoms formed for the first time. This allowed the matter to cool more efficiently, and gravity started to dominate over pressure. With no resistance from pressure, large-scale structures started to form. The wrinkles due to the bubbles of matter created by those acoustic waves are visible today as the baryon acoustic oscillations. The size of the oscillations is determined by the properties of the early Universe and its components: the baryonic matter, the dark matter, and the dark energy. Thus, they can be used to constrain the properties of these components. The imprint of the BAO is left in overdensity peaks at $r_d \sim$  100 Mpc in the two-point statistics of the matter field. Basically, these oscillatory features occur on relatively large scales, which are still predominantly in the linear regime; it is therefore expected that BAOs should also be seen in the galaxy distribution \citep{1999MNRAS.304..851M,2005Natur.435..629S, 2005APh....24..334W,2005ApJ...633..575S,2007ApJ...664..660E}.

The same BAO feature is also present in other clustering measures e.g. the three-point correlation function, or its Fourier image, the bispectrum. In general, the bispectrum is more difficult to model compared to the power spectrum. Recent studies showed that the perturbation theory based models fail to accurately predict bispectrum measurements at surprisingly large scales \citep{ 2016PhRvD..93h3517L,2021PhRvD.104l3551B,2022PhRvD.105f3512I,2022PhRvD.106d3530P,2022MNRAS.512.4961A,2023JCAP...01..031R,2023arXiv230711090K,2023JCAP...08..066G}. Several recent works managed to measure the distance scale from the BAO feature in the bispectrum despite these difficulties \citep{2018MNRAS.478.4500P, 2022PhRvD.105d3517P}. A number of recent works \citep{2015MNRAS.454.4142S,2017MNRAS.468.1070S,2018MNRAS.478.4500P} measured the BAO peak also in the galaxy three-point statistics. More works have studied different aspects of modeling the full bispectrum \citep{2015PhRvD..91l3518B, 2017MNRAS.471.1581B,2017MNRAS.465.1757G,2018MNRAS.477.1604G,2019MNRAS.484L..29G,2022PhRvD.105d3517P,2019MNRAS.484..364S,2020MNRAS.497.1684S,2021MNRAS.501.2862S,2023MNRAS.523.3133S,2023MNRAS.524.1651S} and by comparing theoretical models of increasing difficulty with simulations \citep{2020JCAP...03..056O, 2019MNRAS.482.4883C, 2018JCAP...10..019B}. 

When it comes to the BAO peak position, studies tend to focus on two-point statistics \citep{2012MNRAS.427.3435A,2014MNRAS.441...24A,2016MNRAS.457.1770C,2016MNRAS.460.4210G,2017MNRAS.464.1168R,2017MNRAS.464.3409B}. This is justified by the fact that the reconstruction of the galaxy field \citep{2007ApJ...664..660E,2009PhRvD..80l3501N} moves some of the information from the three-point statistics – e.g. the three-point correlation function or the bispectrum – into the two-point statistics. However, measuring the BAO peak position from the bispectrum directly – or the combination of the non-reconstructed power spectrum and bispectrum – is still useful as the effect of the reconstruction on the information content of the higher-order statistics is, at the moment, not completely clear \citep{2021MNRAS.505..628S}.

In this work, we explore the possibility of isolating the BAO feature in the galaxy bispectrum and modeling it with a simple model supplemented by enough nuisance parameters to account for inaccuracies in modeling its overall shape. The BAO feature in the Fourier space extends to high wavenumbers, but the quantity of interest is not the exact profile of the feature as a function of a wavenumber but its frequency, which is set by large-scale physics. The hope is that, if the BAO feature in the bispectrum is isolated, it will be possible to model it effectively even if the modeling of the full bispectrum is difficult. Our results are encouraging and suggest that surprisingly simple models of the BAO feature in the bispectrum are able to reproduce the BAO frequency with minimal bias. This is not surprising as the BAO signal is imprinted by linear physics in the early Universe. While the exact amount of power at high wavenumbers is difficult to theoretically compute, the quantity of interest in the BAO analysis is the beat frequency of the feature, and this quantity turns out to be more robust.

We use a simple, linear-theory-based model to describe the BAO feature in the bispectrum, in which the effect of nonlinear evolution is modeled by a wavenumber-dependent Gaussian damping term. (see section~\ref{sec:modeling}). We use \texttt{GLAM} simulations (see section~\ref{sec:glam}) to validate this model. These simulations were run in pairs with and without the BAO feature in the initial conditions but otherwise identical. These simulations provide us with the ``empirical'' data for the smooth and wiggly bispectra.  When fitting our model to simulations, we find that the extracted BAO scale is unbiased even when the fiducial template for the BAO is constructed in a slightly offset cosmological model (see section~\ref{sec:fits}). The recovered values of $\alpha$ are within 0.3 per cent of the true values even when the BAO feature in the fitting template deviates from the true value by as much as 3 per cent.

\section{BAO Signal in Simulations}
\label{sec:glam} % used for referring to this section from elsewhere

\subsection{\glam\ simulations}

\begin{figure}
    \includegraphics[width = \columnwidth]{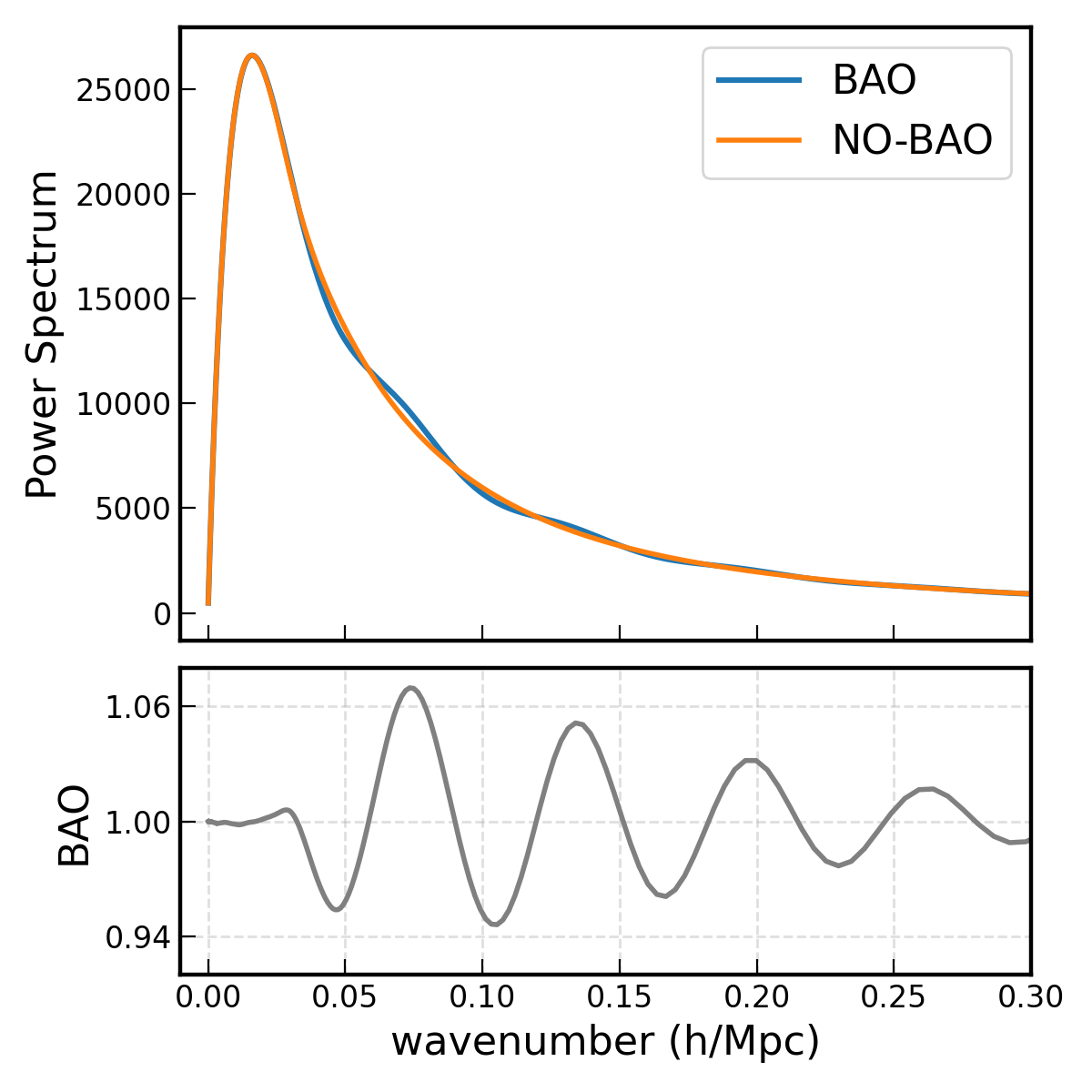}
    \caption{Initial power spectrum of \glam\  \textit{withBAO} and \textit{noBAO} simulations and it's BAO signature. }
    \label{fig:glam_ini}
\end{figure}

We use \texttt{GLAM-PMILL} simulations \citep{CesarGLAM} to test methods for BAO feature separation and theoretical modeling. The \textsc{GLAM-PMILL} catalogs used in this work are publicly available in the \textsc{Skies \& Universes} website\footnote{\url{https://www.skiesanduniverses.org/Products/MockCatalogues/GLAMDESILRG/}}.

The \texttt{GLAM} simulations \citep{2018MNRAS.478.4602K} are $N$-body cosmological simulations that, in the case of the PMILL series, follow the evolution of $2,000^3$ dark matter particles. Each particle has a mass of $1.065\times10^{10}\,h^{-1}\mathrm{M}_\odot$ in a comoving periodic box of size $L=1\,h^{-1}\mathrm{Gpc}$, with $N_{\mathrm{s}}=136$ time-steps and a mesh of $N_{\mathrm{g}}=4,000$ cells per side. This results in a spatial resolution of $0.250\,h^{-1}\mathrm{Mpc}$. 

The simulations were produced in a flat \textLambda CDM cosmology, adopting the cosmological parameters assumed in the Planck Millennium simulation \citep[PMILL;][]{Pmill_cosmo}. PMILL uses the best-fitting parameters from the first Planck 2013 data release \citep{2014A&A...571A..16P}. Each instance of the simulation was run with two initial conditions, starting at $z_\mathrm{ini}=100$: one with the BAO signal imprinted in the initial conditions, and one without. The pairs of simulations are otherwise identical, i.e. they had matching random phases in the initial conditions. The \texttt{GLAM-PMILL} simulations used in this work consist of 1,000 realizations with and another 1,000 without BAO. 
Figure~\ref{fig:glam_ini} shows the power spectrum used for the initial conditions of the \glam\ simulations. The initial power for the pair of simulations was identical, except the ``withBAO'' instances had periodic BAO wiggles in them, and the ``noBAO'' instances didn't have the feature.

The \texttt{GLAM} simulations are only capable of correctly resolving distinct halos (not subhalos), with virial masses greater than $\sim10^{12}\,h^{-1}\mathrm{M}_\odot$ for the case of the \texttt{GLAM-PMILL} suite. The resulting dark matter halos were populated by galaxies using GALFORM semi-analytical model of galaxy formation \citep{CesarGLAM}, but in this work we only use the halo catalogs, both for BAO and noBAO realizations.

We will treat the measurements from these simulations as the ground truth for the rest of our work. The theoretical models will be evaluated based on how well they recover the measurements of \glam\ simulations.

\subsection{BAO in the power spectrum}

We measure the power spectrum from both the \textit{withBAO} and \textit{noBAO} simulations, $\Pw$ and $\Ps$ respectively (see Appendix~\ref{app:ps_measurements} for the details). The power spectrum measurements are averaged in bins of width $\Delta k = 0.01h\ \Mpc^{-1}$ starting from $k_\mathrm{min} = 0.005h\ \Mpc^{-1}$ up to $k_\mathrm{max} = 0.3h\ \Mpc^{-1}$, resulting in 30 bins. This wavenumber range contains most of the linear and semi-linear modes of galaxy overdensity. There is very little BAO signal above our maximum wavenumber. 

\begin{figure}
    \includegraphics[width = \columnwidth]{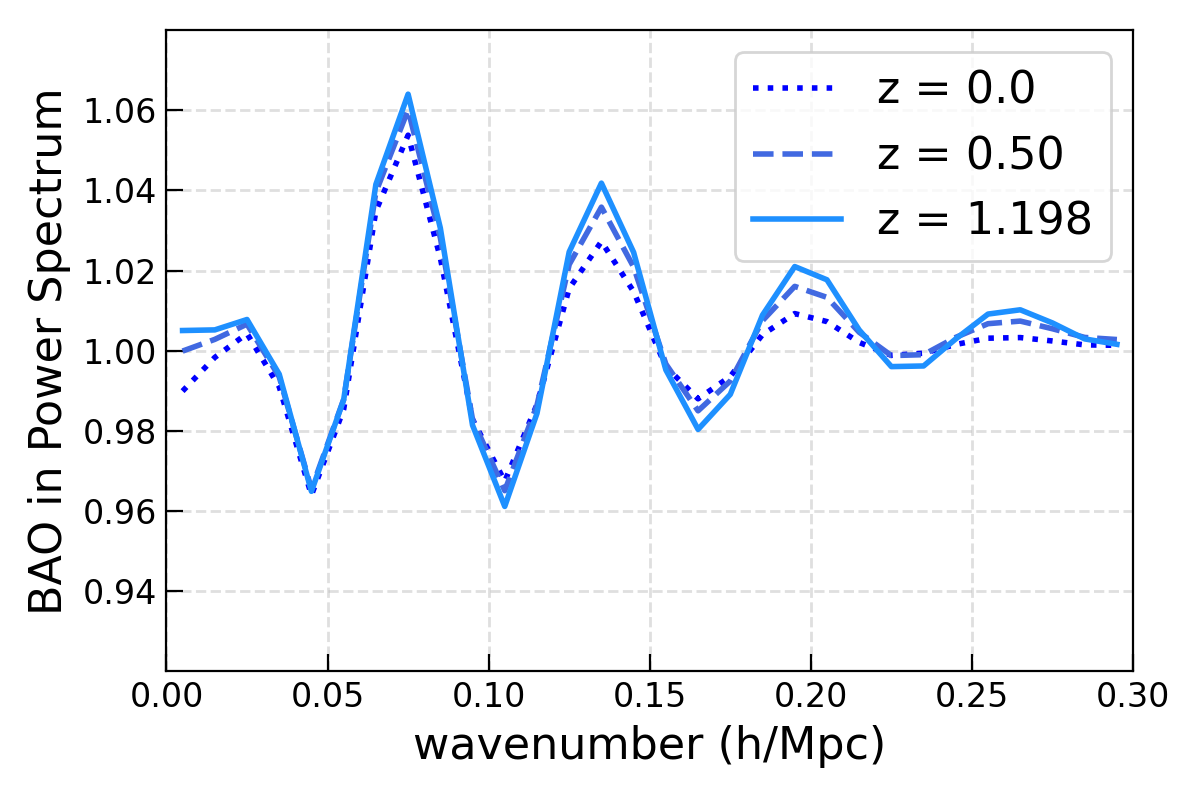}
    \caption{BAO signature in the power spectrum of \texttt{GLAM} simulations at different redshifts.}
    \label{fig:glam_ps_bao}
\end{figure}

Figure~\ref{fig:glam_ps_bao} shows the ratio of $\Pw$ to $\Ps$ measured from \glam\ at different redshifts. This is by definition the BAO signal in the power spectrum. The BAO signal looks exactly as expected. The amplitude of wiggles decays with redshift. The decay is wavenumber-dependent. BAO in high wavenumber modes decays faster with redshift as expected.

\subsection{BAO in the bispectrum}
\label{sec:red_bk}

\begin{figure}
    \includegraphics[width = \columnwidth]{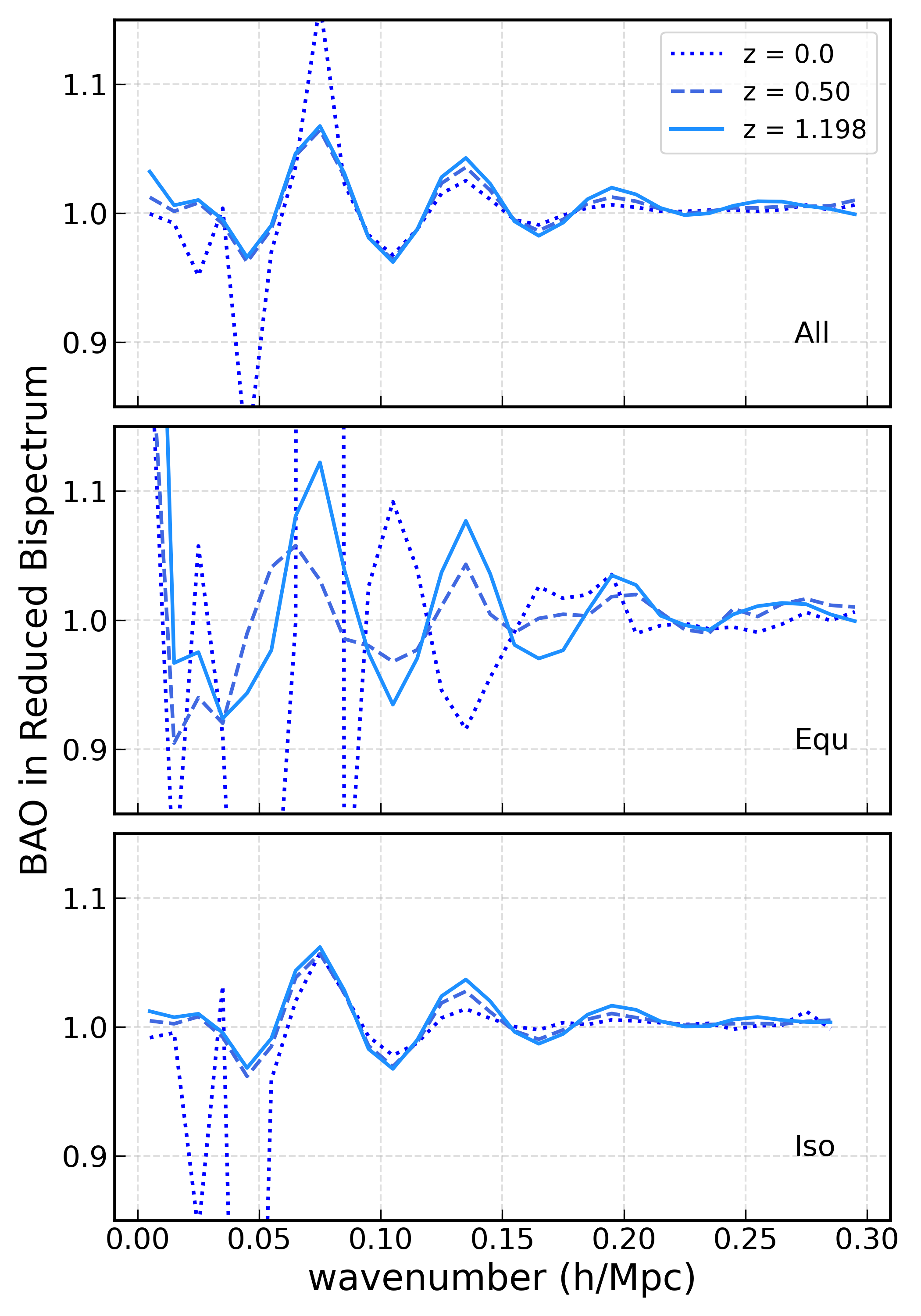}
    \caption{BAO signature in the bispectrum of \texttt{GLAM} simulations at different redshifts. Labels All (top panel), Equ (middle panel), and Iso (lower panel) refer to averages over all, equilateral, and isosceles configurations respectively. }
    \label{fig:glam_bs_bao}
\end{figure}

We define the BAO feature in the bispectrum similarly to that of the power spectrum. We measure the angularly averaged bispectrum - $B(k_1, k_2, k_3)$ - from \textit{withBAO} and \textit{noBAO} \texttt{GLAM} simulations, $\Bw$ and $\Bs$ respectively (see Appendix~\ref{app:ps_measurements} for the details). We bin each wavevector in 30 bins similar to how we did it for the power spectrum. This results in a total of 2600 bins for the $(k_1, k_2, k_3)$ triplet. To map the 3D configuration space of triangles into a 1D index for analysis, we employ a “triangle index” that orders the triangles so that $k_1 \leq k_2\leq k_3$. This ensures each unique triplet is only counted once without duplication since the bispectrum is invariant to any permutation of the indices.

% To map the 3D configuration space of triangles into a 1D index for analysis, we employ a “triangle index” that systematically orders the triangles. We define the index $i$ as,
% \begin{equation}
%     i(k_1, k_2, k_3) = \sum_{j=1}^{3} \frac{k_j - k_{min}}{dk}N^{3-j} + 1
% \end{equation},
% where $k_j$ are the three wavenumbers, and $k_{min}$ and $k_{max}$ are the minimum and maximum wavenumber values in the data. The indexing starts with the smallest, most equilateral triangle then increments one $k_i$ at a time, exploring triangles with one increasingly longer side. When a $k_i$ reaches the maximum wavenumber, it is reset to the minimum and the next $k_j$ is incremented instead. This continues until all configurations are indexed. Thus, in the resulting dataset ordered by the triangle index $i$, nearby index values correspond to triangles that differ in the length of only one side, enabling analysis of how the bispectrum changes with overall shape and scale of the wavenumber triangles. The index ordering does not directly sort by perimeter or other metrics, but serves as a systematic mapping from the 3D triangle space to a 1D dataset.

Figure~\ref{fig:glam_bs_bao} shows the ratio of $\Bw$ to $\Bs$ at different redshifts. For better visualization, the measurements are averaged over $k_2$ and $k_3$ for different triangular configurations. We look at three different averages. In $B^\mathrm{all}(k)$ the value of $k_1$ is kept within a certain bin while all other values of $k_2$ and $k_3$ are averaged over. In $B^\mathrm{iso}(k)$ we average over bispectra that have the value of $k_1$ within a certain bin and $k_2 = k_3$. In $B^\mathrm{equ}(k)$ we average over all bispectra that have all three wavenumbers within a certain bin.

\begin{align}
    % \hat{B}(k) &= \displaystyle\int\!\! B(k, k_2, k_3) \mathrm{d}k_2\mathrm{d}k_3 \\
    B^\mathrm{all}(k) &= \displaystyle\sum_{k_2,k_3} \frac{B(k,k_2,k_3)}{N_k}
    \label{eq:red_all}\\
    B^\mathrm{iso}(k) &= \displaystyle\sum_{k_2,k_3} \frac{B(k,k_2,k_3)_{k\neq k_2=k_3}}{N_{k\neq k_2=k_3}}
    \label{eq:red_iso}\\
    B^\mathrm{equ}(k) &= \displaystyle\sum_{k_2,k_3}\frac{B(k,k_2,k_3)_{k=k_2=k_3}}{N_{k=k_2=k_3}},
    \label{eq:red_equ}
\end{align}
where, $N_k$, $N_{k\neq k_2=k_3}$ and $N_{k=k_2=k_3}$ are the number of triangles that fall into the relevant configurations. We only do this to make the comprehansion of the plots easier. All our analysis are performed on the original (pre-averaged) 2600 bispectrum measurements.

Even though this averaging is not optimal for the retention of the BAO feature, it is still clearly visible in all configurations. The equilateral configuration has the most noise since each point in the $k$ axis corresponds only to a single  equilateral triangle. In the bispectrum, the BAO feature looks qualitatively similar. It is a decaying oscillation which is more pronounced at higher redshifts.

\section{BAO modeling}
\label{sec:modeling}
\subsection{Power spectrum}

To model the BAO in the power spectrum we start with the linear model $\Plin(k)$ that we compute using computer code \camb\ \citep{Lewis:2002ah} in a fiducial cosmological model. We split this power spectrum into BAO and smooth parts, $\Plin^\mathrm{BAO}$ and $\Plin^\mathrm{s}$ respectively, using \href{https://github.com/cosmodesi/cosmoprimo}{\texttt{cosmoprimo}} package.
\texttt{cosmoprimo} has many options for performing this split. We use the option based on the work in \cite{2018PhDT.......146W} and \cite{2010JCAP...07..022H}, but we verified that this choice does not affect our results. We reduce the BAO feature by damping it with a Gaussian term $D(k, \Sigma) = \exp(-\Sigma^2k^2/2)$ so that our final power spectrum template is
\begin{equation}
    P^\mathrm{w}(k;\alpha, \Sigma) = \Plin^\mathrm{s}(k)[(\Plin^\mathrm{BAO}(\alpha k) - 1)D(k, \Sigma) + 1]
\end{equation}
The dilation parameter $\alpha$ is only applied to the BAO wiggles. 
The superscript w denotes the fact that the template has BAO wiggles in it. We use just the smooth part of the linear power spectrum as the no-wiggle template.
\begin{equation}
    P^\mathrm{nw}(k) = \Plin^\mathrm{s}(k)
\end{equation}
Formally, this is a limit $\Sigma\rightarrow 0$ of the wiggled template.

\subsection{Bispectrum}

Our bispectrum BAO model is inspired by the leading order perturbation theory with linear local bias given by \cite{2000ApJ...544..597S}.
\begin{align}\label{bk_eqn}
    B(\vect{k}_1,\vect{k}_2,\vect{k}_3) &= 2 \vect{Z}_1(\mu_1)\vect{Z}_1(\mu_2)\vect{Z}_2P(\alpha k_1)P(\alpha k_2) + \\ \nonumber
    & \text{cyclic terms} + [P(k_1) + P(k_2) + P(k_3)]S_1 + S_0 ,
\end{align} 
\\
where,

\begin{equation}\label{Z1_eqn}
    \vect{Z}_1(\mu) = (b_1+f\mu_1^2),
\end{equation}

\begin{align}
 \vect{Z}_2 &= \Bigg[\frac{b_2}{2} + b_1 F_2(\vect{k}_1,\vect{k}_2) + f\mu_3^2 G_2(\vect{k}_1,\vect{k}_2)\\ \nonumber
    &- \frac{f\mu_3k_3}{2}\bigg(\frac{\mu_1}{k_1}(b_1+f\mu_2^2) + \frac{\mu_2}{k_2}(b_1+f\mu_1^2)\bigg)\Bigg],
\end{align}

$\mu_i = \vect{k}_i\cdot\vect{\hat{z}}/k_i$ with $\vect{k}_1 + \vect{k}_2 + \vect{k}_3 = 0$   to account for the triangular condition.

\begin{equation}\label{F2_eqn}
    F_2(\vect{k}_1,\vect{k}_2) = \frac{5}{7} + \frac{\vect{k}_1\cdot\vect{k}_2}{2k_1 k_2}\bigg(\frac{k_1}{k_2}+\frac{k_2}{k_1}\bigg) + \frac{2}{7}\bigg(\frac{\vect{k}_1\cdot\vect{k}_2}{k_1 k_2}\bigg)^2,
\end{equation}
\\
and

\begin{equation}\label{G2_eqn}
    G_2(\vect{k}_1,\vect{k}_2) = \frac{3}{7} + \frac{\vect{k}_1\cdot\vect{k}_2}{2k_1 k_2}\bigg(\frac{k_1}{k_2}+\frac{k_2}{k_1}\bigg) + \frac{4}{7}\bigg(\frac{\vect{k}_1\cdot\vect{k}_2}{k_1 k_2}\bigg)^2,
\end{equation}

The cyclic terms can be derived by replacing indices 1 and 2 in the first term with 2 and 3, and 1 and 3 respectively. $f$, $b_1$, $b_2$, $S_0$ and $S_1$ are free parameters of the fit.

\cite{2017MNRAS.467..928G} showed that most of the information is in the azimuthal averages of the first three even multipoles with
$m = 0$ in the multipole expansion containing most of the information relevant to the derivation of cosmological constraints.
In this work, we only consider the angularly averaged bispectrum (or the monopole) defined as
\begin{equation}\label{B0_eqn}
B_0(k_1,k_2,k_3) =  \int\limits_{-1}^{+1} d(\cos{\theta_1}) \int\limits_{0}^{2\pi} d\phi B(k_1,k_2,k_3,\theta_1,\phi).
\end{equation}
We integrate the five-dimensional bispectrum numerically to obtain the three-dimensional bispectrum monopole.

The model depends on wavevector lengths $k_1$, $k_2$ and $k_3$, a dilation parameter $\alpha$ and nuisance parameters $b_1$, $b_2$, and $f$. In perturbation theory, the last three parameters have a physical meaning (e.g. $f$ is interpreted as a growth rate of structure). We treat these parameters as true nuisance parameters. Their only purpose is to give us a reasonable range of bispectrum shapes to separate the BAO feature. We do not assign to them any physical meaning. 

We compute $B$ by using $P^\mathrm{w}$ and $B^\mathrm{s}$ by using the no-wiggled template $\Ps$. The ratio $B/B^\mathrm{s}$ then gives a model for the BAO signature in the bispectrum.

In the end, the model depends on parameters $\alpha$, $b_1$, $b_2$, $f$, $S_0$, $S_1$, and $\Sigma$.

This single universal scaling model is not accurate at the sub-percent precision \citep{2023arXiv230716498K} but we ignore the higher order effects in this work since they do not really affect any of our main conclusions.

\section{BAO fitting}
\label{sec:fits}

We fit the model described in section~\ref{sec:modeling} to the \texttt{GLAM} measurements described in section~\ref{sec:glam} by running a Monte Carlo Markov Chain (MCMC) using the \emcee\ Python package \citep{2013PASP..125..306F}. The MCMC was run with 70 walkers for 2000 steps, discarding the first 800 steps as burn-in to ensure convergence. The chain is run over $\alpha$ and 6 nuisance parameters ($f$, $b_1$, $b_2$, $S_0$, $S_1$, $\Sigma$). The first three parameters in that list control the relative amplitude at different wavenumber triplets, the next two provide wavenumber-dependent and wavenumber-independent additive terms to remove the effect of shot noise, and the last parameter controls how pronounced the BAO feature is.
The parameter priors are wide enough to encompass all areas of high likelihood. We find the best-fit values by minimizing 
\begin{equation}
\label{eq:chi2}
    \chi^2 = \displaystyle\sum_{ij} X_i W_{ij} X_j,
\end{equation}
where $X$ is a vector of measurements (either $P_\mathrm{w}/P_\mathrm{s}$ at different wavenumber or $B_\mathrm{w}/B_\mathrm{s}$ at different wavenumber triplets) and $W_{ij}$ are weights that upweight or downweight a relative contribution from different wavenumbers to the overall fit. The most optimal weights are the ones given by the inverse covariance of our measurements. Unfortunately, we don't have a reliable way of estimating the inverse covariance. We use the inverse of a sample variance over 1,000 \texttt{GLAM} mocks as our weights. We can not use the posterior of our likelihood to approximate the true constraining power of the BAO signal in the bispectrum since the sample covariance computed from a small number of mocks is a very noisy estimator of a true covariance and the Hartlap correction factor \citep{2007A&A...464..399H}, commonly used to unbias the inverse of the covariance matrix, was not applied in this case due to the insufficient number of realizations. The minimum of the Eq.~\ref{eq:chi2} is still an unbiased estimator and we can use it to estimate the systematic offsets in our fits. Our measurements are computed from a very large cumulative volume of 1,000$h^{-1}$ Gpc and we expect the uncertainty in the systematic offset to be negligible.

$\alpha$ is the only parameter of interest to us, the other six are nuisance parameters that we marginalize over. We have another hidden degree of freedom in this fit, which is the fiducial cosmology that we choose to produce the original $\Plin$ template. We are hoping that no matter which template we start with, it can be made to fit the measurements by an appropriate rescaling of the oscillation frequency with $\alpha$ and damping the amplitude of oscillations at high wavenumbers with $\Sigma$. The dilation parameter is expected to be
\begin{equation}
    \alpha = \frac{\rd^\mathrm{tmp}}{\rd}.
\end{equation}
We do not have scaling with $D_V$ like in Eq.~\eqref{alphadefinition}. This is because 
we are measuring our bispectra in cubic boxes with known physical size and we are not affected by Alcock-Patczynski scaling. We will show the results of fitting to the redshift $z=0.5$, but we also verified that the same conclusions hold for redshifts $z=0$ and $z=1.198$.

\begin{figure*}
    \includegraphics[width = \textwidth]{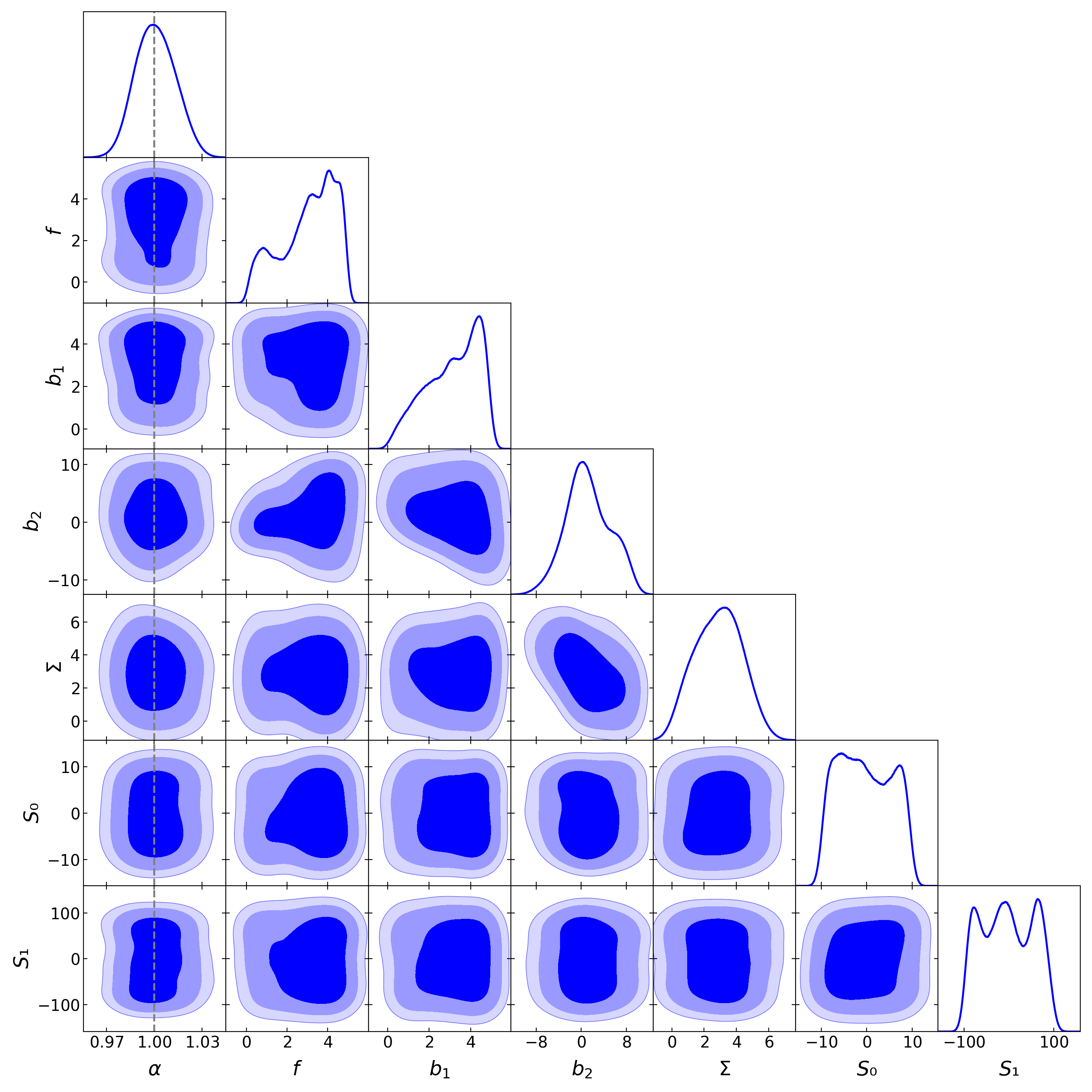}
    \caption{Equipotential contours of the MCMC posterior from fitting the BAO in \texttt{GLAM} bispectrum at $z = 0.5$. The grey dashed line shows the expected value of $\alpha = 1$.}
    \label{fig:contour_p7_a1camb}
\end{figure*}

\begin{figure}
    \includegraphics[width = \columnwidth]{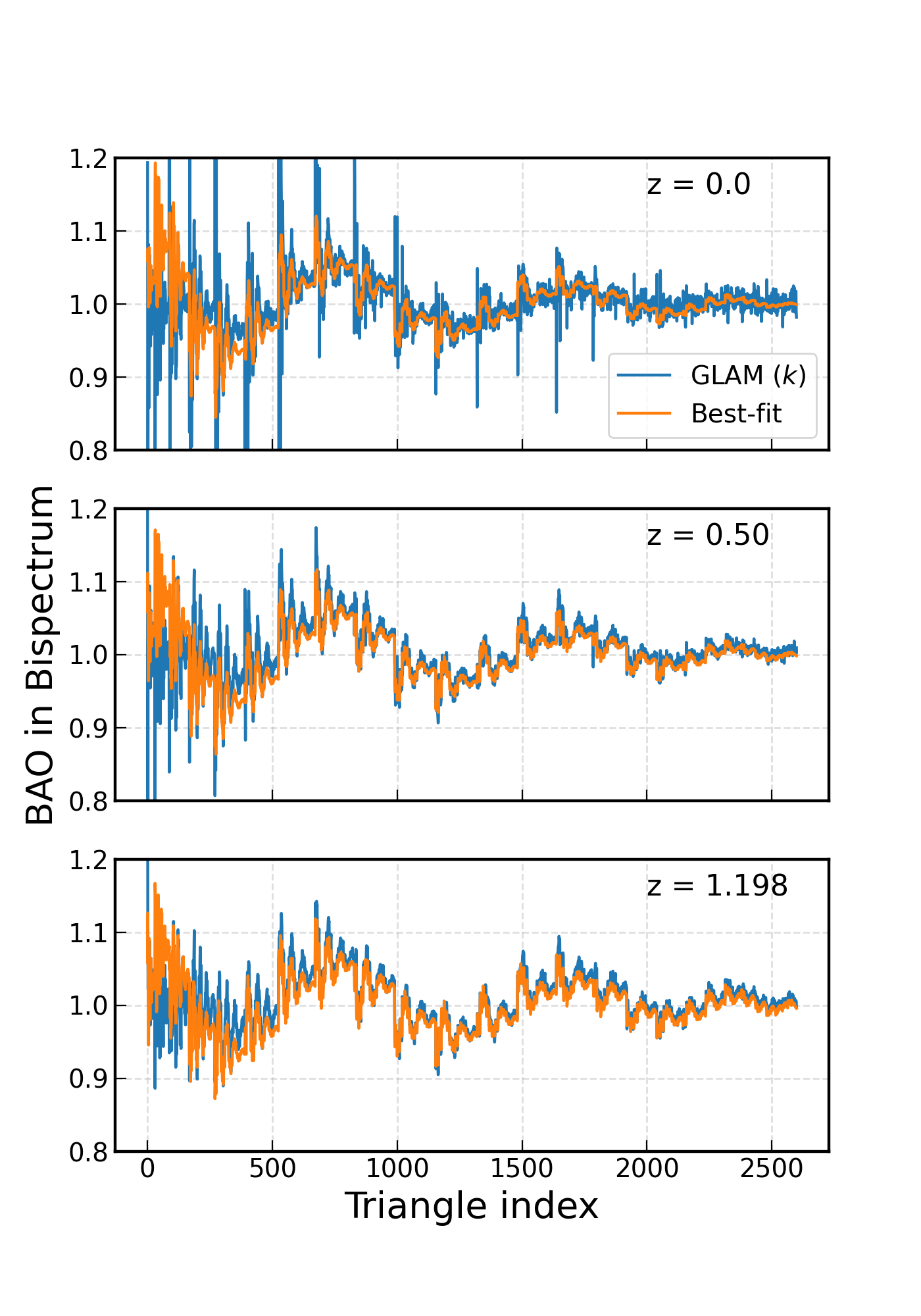}
    \caption{BAO signature in the bispectrum as a function of triangle index (see definition in Sec~\ref{sec:red_bk}) at different redshifts. Measured bispectrum BAOs from the \texttt{GLAM} simulations (the blue line) is compared to our best-fit model (the orange line).}
    \label{fig:bk_a1}
\end{figure}

\begin{figure}
    \includegraphics[width = \columnwidth]{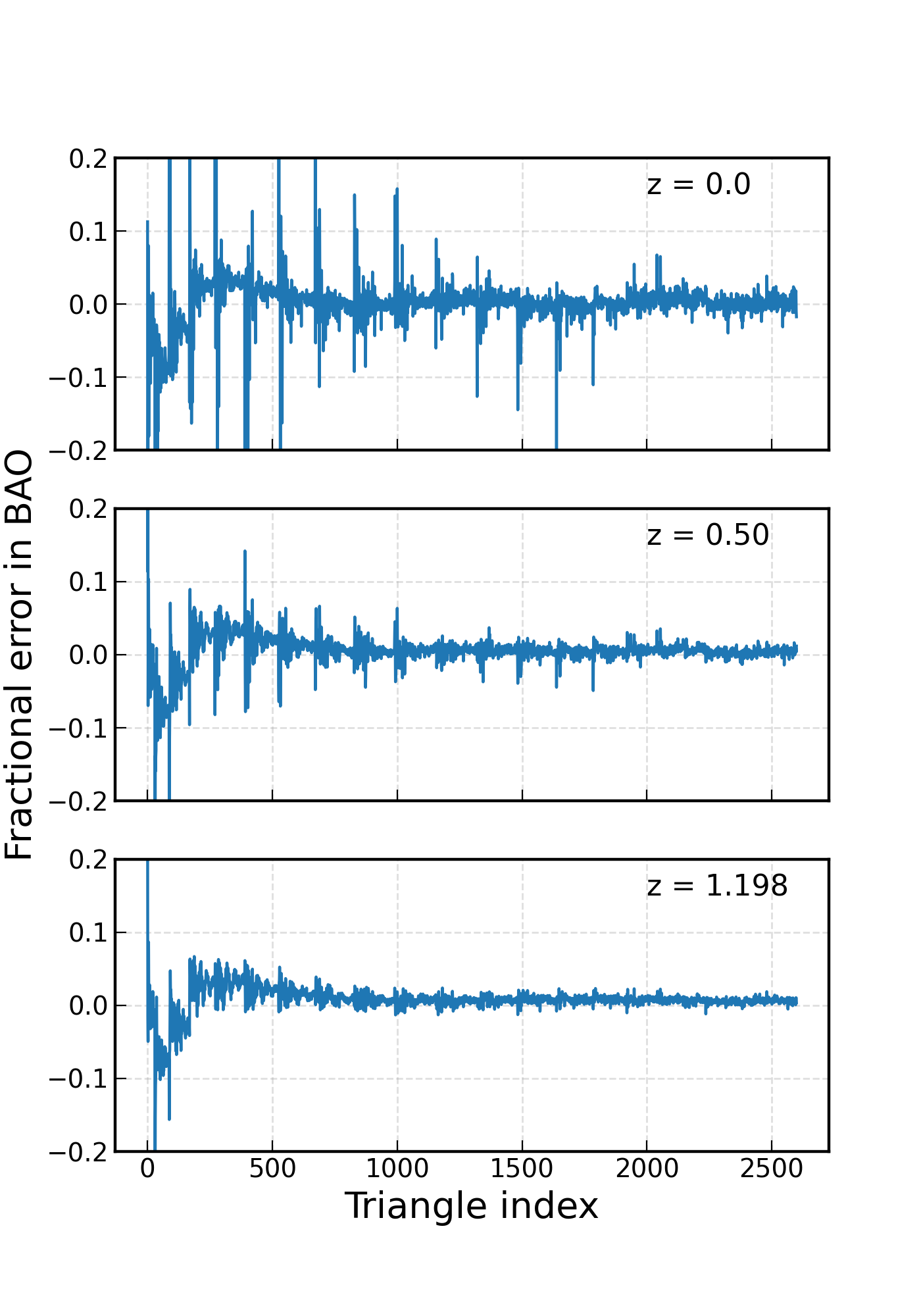}
    \caption{Fractional deviation between the measured bispectrum BAOs from the \texttt{GLAM} simulations and our best-fit model as a function of triangle index (see definition in Sec~\ref{sec:red_bk}) at different redshifts.}
    \label{fig:frac_bk_a1}
\end{figure}

\begin{figure}
    \includegraphics[width = \columnwidth]{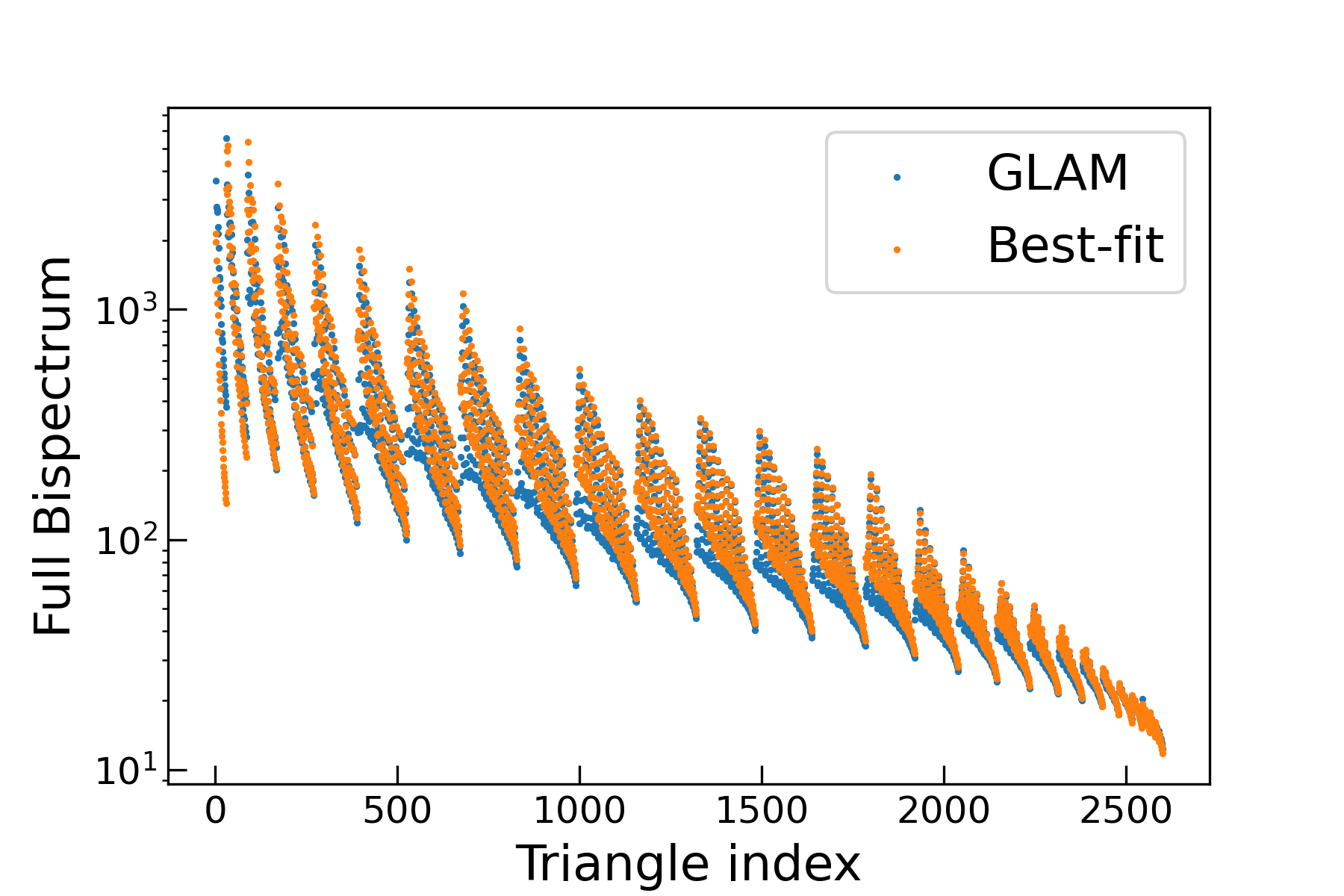}
    \caption{Full bispectrum as a function of triangle index (see definition in Sec~\ref{sec:red_bk}) at redshift 0.5. Measured bispectrum from the \texttt{GLAM} simulations (the blue line) is compared to our best-fit model (the orange line).}
    \label{fig:fullbk_glam_camb}
\end{figure}

\begin{figure*}
    \includegraphics[width = \textwidth]{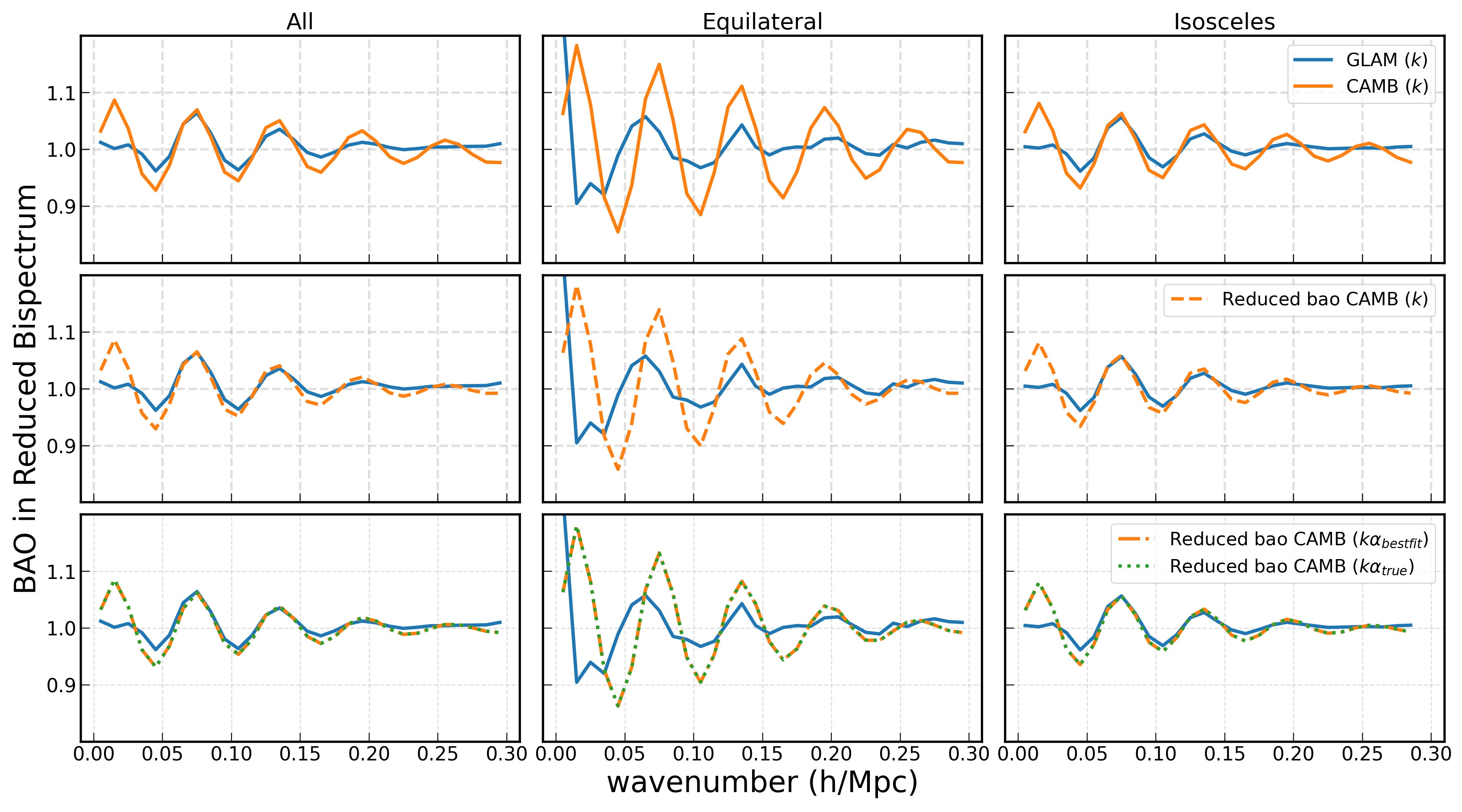}
    \caption{Measured BAO in the bispectrum compared to the model. The columns show bispectrum averaged over all (left), equilateral (middle), and isosceles (right) configurations. The blue lines denote the measurements and are identical along the columns. The solid orange lines in the top row denote the model computed based on a \texttt{CAMB} linear power spectrum. The dashed orange lines in the middle row denote the model with the BAO damping applied, which reduces the amplitude of the BAO wiggles. The dot-dashed yellow line and the dotted green line in the bottom row denote the model stretched by the best fit and true $\alpha$ values respectively.}
    \label{fig:red_bk_a1}
\end{figure*}

\begin{figure}
    \includegraphics[width = \columnwidth]{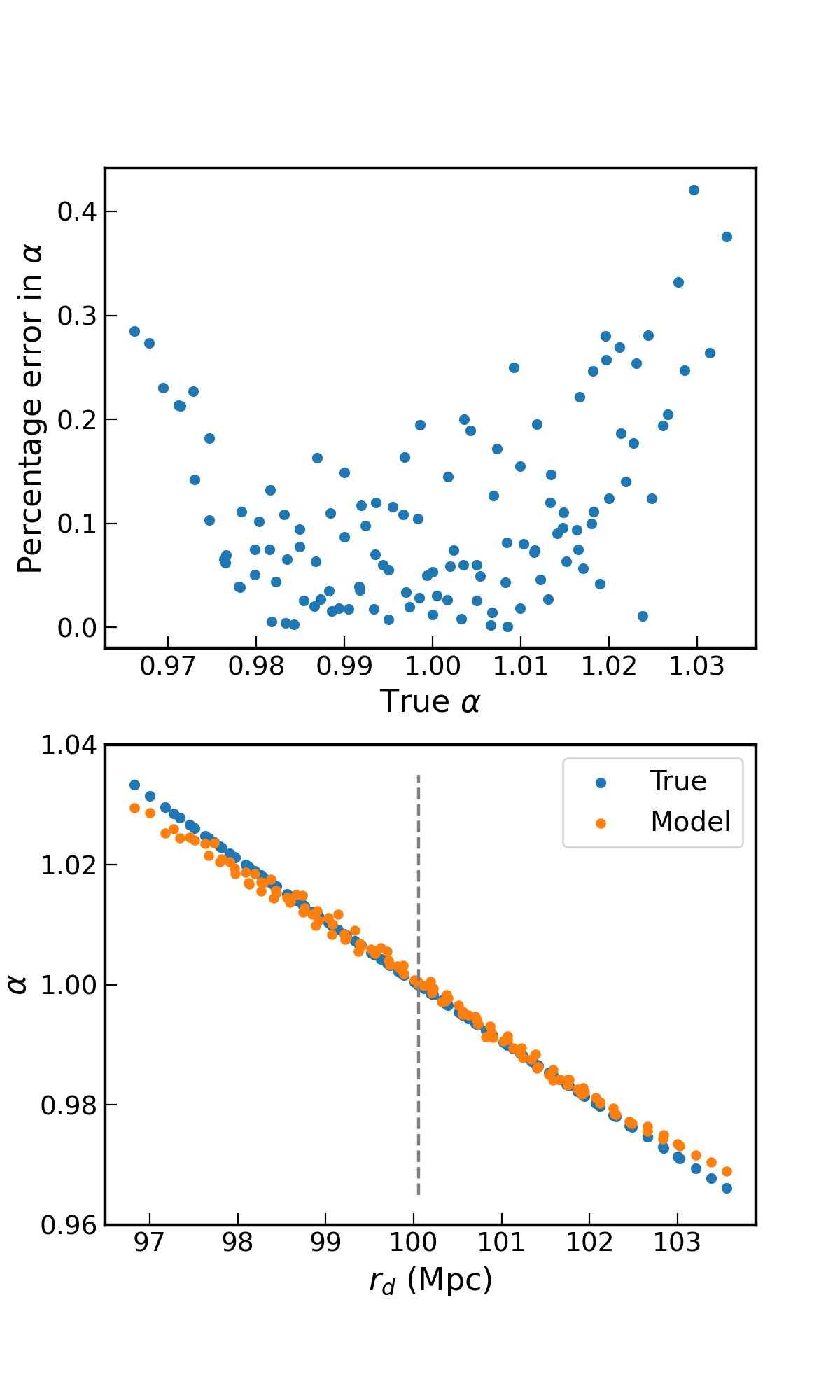}
    \caption{Percentage systematic error in the recovered value of $\alpha$ as a function of the expected alpha. }
    \label{fig:systematic}
\end{figure}

\subsection{Aligned Template}

We first check what happens when the template is computed in the actual cosmology of \texttt{GLAM} simulations. In this case, the BAOs in the template and the measurements are aligned and the expected value of $\alpha$ is one.
The results from running our MCMC chains for the aligned template for the redshift of $z=0.5$ are shown in Figure~\ref{fig:contour_p7_a1camb}. The results for other redshifts look qualitatively similar. The likelihood contours are reasonably close to Gaussian. While some of the nuisance parameters are correlated with each other, the $\alphav$ parameter is uncorrelated with any of the nuisance parameters, suggesting that its maximum likelihood values are weakly affected by possible posterior volume effects. The $f$ parameter is allowed to take on negative values, under the standard interpretation of its meaning this would be unphysical. We, however, do not interpret $f$ as a growth rate of structure but rather use it as a pure nuisance parameter that is used to help us span the range of possible shapes for the smooth part of the bispectrum, so this is not a problem for our analysis. We checked that restricting $f$ to have only positive values does not affect constraints on $\alpha$, which is not surprising since $f$ and $\alpha$ are very weakly correlated. Our recovered value of $\alpha$ is consistent with unity to 0.03 percent precision. It's worth noting that since we do not have reliable estimates of the covariance matrix, the widths of our MCMC chains can not be interpreted as errors in the measurement. We do not have at present a good way of estimating the variance in our measurements but we expect it to be subdominant to the systematic error from the fit since we are fitting the mean measurement of 1,000 one cubic Gigaparsec mocks.

Figure~ \ref{fig:bk_a1} shows the bispectrum BAOs from \texttt{GLAM} measurements at redshifts $0.0$, $0.5$, and $1.198$, and their corresponding best-fit models evaluated at the best values derived from MCMC chains. The models match the measurements well and by visual inspection successfully recover all features in the measurements.Figure~\ref{fig:frac_bk_a1} shows the same information but now as a ratio of the predicted and measured bispectra. At $z=0.0$ there are some bispectrum triplets that have a large fractional deviation between the measurement and the theory. This is likely the failure of the simple model adopted in our work which does capture the overall frequency of the BAO oscillations well but fails to predict the amplitude of the BAO for some triplets.

Figure~\ref{fig:fullbk_glam_camb} shows a similar best fit to the entire bispectrum (not the BAO only like in the previous case). As expected the simple model can not fit the full shape of the bispectrum well, there are obvious and irreconcilable offsets between the measured and theoretical bispectra at a wide range of wavenumbers and triangular configurations.

To aid visual inspection we also plot the measured and best-fit model bispectra for averaged, isosceles, and equilateral triangles as defined by Eqs.~\eqref{eq:red_all}-\eqref{eq:red_equ}. Figure~\ref{fig:red_bk_a1} shows this comparison for the bispectrum measured at $z = 0.5$. The top panel shows the \texttt{GLAM} BAOs along with the model computed for the original templates derived from \texttt{CAMB}. The middle row shows the effect of the damping parameter $\Sigma$ on the model for the best-fit $\Sigma$ from the MCMC chains. The bottom row shows the result of stretching it by the best-fit value of $\alpha$. Since our best-fit is very close to one, the best-fit BAO template is difficult to distinguish from the ``true'' BAO template (evaluated at $\alphav = 1$), which we also plot on the same figures for comparison. Visual inspection confirms that the best-fit template is indeed a good fit for the harmonic feature imprinted in the measurements. The template seems to be offset for the equilateral triangles. This can be explained by high noise in this specific measurement. The averaged and isosceles measurements are averages over many triangular configurations, while the equilateral measurement is averaged over very few triangular configurations.

\subsection{Systematic effects due to offset template}

We now check whether the model performs equally well when the template is computed in a different cosmology from the \texttt{GLAM} so that the position of the BAO peak is offset. To perform this test we generate templates in 100 different cosmologies around \texttt{GLAM} cosmology in a range $ 0.277 \leq \Omega_m \leq 0.327$. We then perform the fitting as before. Figure \ref{fig:systematic} shows the results of this exercise. Each point on the top panel corresponds to a template used in the fits. The horizontal axis shows the expected value of $\alpha$ and the vertical axis shows the percentage offset of the best fit from that expectation. In the ideal case, we want the points to be as close as possible to zero. The bottom panel shows the same data from a different point of view. The points are again individual templates, the horizontal axis is the value of $\rd$ in the template and $\alpha$ is the amount of dilation needed to bring the template to be identical to the true BAO signal. Blue dots show our expectations and the orange dots show the actual results. We expected the systematic offset between the two to increase as the offset between the template and the simulations increases, which seems to be the case. The systematic offsets are mostly within 0.3 per cent as long as the frequency of the BAO in the template does not differ by more than 3 per cent from the frequency in the data.

\section{Conclusions}

We explored the possibility of modeling the BAO feature in the galaxy bispectrum. Even though some of this information is recovered during the reconstruction of the cosmological fields, the measurements of the BAO in the galaxy bispectrum have the potential to strengthen our constraints on key cosmological parameters. In the measured bispectrum the BAO signal is visible at wavenumber as high as $k \sim 0.3h^{-1}$ Mpc at redshifts of $z = 0.5$ and above. At lower redshifts, the BAO signature at higher wavenumber is damped by nonlinear evolution. We showed that even though simple perturbation-based models of bispectrum fail to model it accurately at higher wavenumbers, they work reasonably well when the smooth component is subtracted and only the BAO feature is modeled. We validated this on a suit of \texttt{GLAM} simulations that have been run with and without the BAO feature in the initial conditions. The ratio of the bispectra measured from those simulations is by definition the BAO feature in the bispectrum. Our six-parameter model was able to recover the unbiased value of $\alpha$ parameter when fit to the bispectrum monopole measured from a cumulative volume of $1,000\,h^{-3}\mathrm{Gpc}^3$. In real analysis, we don't know a priori what cosmological model to use when producing BAO templates. It is therefore important that the procedure results in unbiased results for a range of reasonable templates. We checked the robustness of the model by fitting the same measurements with the templates computed in offset cosmologies. By performing similar validation tests we demonstrated that the bias in the recovered values of $\alpha$ is at most 0.3 per cent for deviations up to 3 per cent from the true value of the sound horizon. This is good enough when fitting to currently existing data but may need improvement when future more precise data arrives (e.g. from the DESI and Euclid experiments). Possible avenues for decreasing the bias are computing the BAO template with more sophisticated modeling of nonlinear physics or using simulations for calibrating the templates. 

\section*{Acknowledgements}

We are grateful to Francisco Prada and Anatoly Klypin for their help with \texttt{GLAM} simulations. 

This research used resources of the National Energy Research Scientific Computing Center (NERSC), a U.S. Department of Energy Office of Science User Facility located at Lawrence Berkeley National Laboratory, operated under Contract No. DE-AC02-05CH11231

This research has made use of NASA’s Astrophysics Data System and the arXiv open-access repository of electronic preprints and postprints.

JB, MR, and LS are grateful for support from the US Department of Energy via grants DE-SC0021165 and DE-
SC0011840. JB is partially supported by the NASA ROSES
grant 12-EUCLID12-0004.
JE acknowledges financial support from the Spanish MICINN funding grant PGC2018-101931-B-I00.
We thank Instituto de Astrofísica de Andalucía CSIC for hosting the \texttt{Skies \& Universes} site for cosmological simulation products. GLAM A-series simulations were performed at the DIRAC@Durham facility managed by the Institute for Computational Cosmology on behalf of the STFC DiRAC HPC Facility in the UK. 

%%%%%%%%%%%%%%%%%%%%%%%%%%%%%%%%%%%%%%%%%%%%%%%%%%
\section*{Data Availability}

 Any data product generated during this research will be shared upon request. The \texttt{GLAM} simulations are available for download from \url{https://www.skiesanduniverses.org/Products/MockCatalogues/GLAMDESILRG/}.



\bibliographystyle{mnras}
\bibliography{references} % if your bibtex file is called references.bib

%\begin{thebibliography}{99}
%\end{thebibliography}

%%%%%%%%%%%%%%%%%%%%%%%%%%%%%%%%%%%%%%%%%%%%%%%%%%

%%%%%%%%%%%%%%%%% APPENDICES %%%%%%%%%%%%%%%%%%%%%

\appendix

\section{Measuring Power Spectrum and Bispectrum}
\label{app:ps_measurements}
%%%%%%%%%%%%%%%%%%%%%%%%%%%%%%%%%%%%%%%%%%%%%%%%%%

We divide every simulation cube into a $1024^3$ grid. We assign halos to grid points based on the triangular shaped cloud scheme described in \citep{2016MNRAS.460.3624S} to compute the number of particles in each grid cell - $n_{ijk}$. We then compute the overdensity field 
\begin{equation}
    \delta_{ijk} = \frac{n_{ijk} - \overline{n}}{\overline{n}},
\end{equation}
where $\overline{n}$ is the average number density
\begin{equation}
    \overline{n} = \frac{1}{1024^3}\displaystyle\sum_{ijk} n_{ijk}.
\end{equation}
We then perform a 3D discrete Fourier transform of the overdensity field
\begin{equation}
    \tilde{\delta}_{ijk}= \mathcal{F}_{ijk}^{\ell mn}[\delta_{\ell mn}].
\end{equation}
The $\tilde{\delta}_{ijk}$ numbers can be arranged in a 3D cube so that each measurement corresponds to a certain wavevector $\mathbf{k} = (k_i, k_j, k_k)$ \citep{1983dfti.book.....G, 1985ffc..book.....R,1988ffti.book.....B}.
The average power spectrum within a bin $(k_\mathrm{min}, k_\mathrm{max})$ is computed as the average value of all $\tilde{\delta}_{ijk}\tilde{\delta}^\star_{ijk}$ such that the length $|\mathbf{k}| = \sqrt{k_i^2 + k_2^2 + k_3^2}$ falls within the bin. The average bispectrum within a bin $(k_{i,\mathrm{min}},  k_{i,\mathrm{max}})$, where $i=(1,2,3)$ is similarly computed as the average of all $\tilde{\delta}_{i_1j_1k_1}\tilde{\delta}_{i_2j_2k_2}\tilde{\delta}_{i_3j_3k_3}$ such that wavevectors associated to the three $\tilde{\delta}$ fall within the corresponding bins and their sum is zero.

% Don't change these lines
\bsp	% typesetting comment
\label{lastpage}
\end{document}